\documentclass[conference]{IEEEtran}

\pdfoutput=1

\ifCLASSINFOpdf
\usepackage[pdftex]{graphicx}
\else
\fi

\usepackage{enumerate}
\usepackage[cmex10]{amsmath}
\usepackage{algorithm}
\usepackage[noend]{algpseudocode}

\makeatletter
\def\BState{\State\hskip-\ALG@thistlm}
\makeatother

\usepackage{slashbox}

\begin{document}

%
\title{A Systematic Scheme for Measuring the Performance of the Display-Camera Channel}

\author{\IEEEauthorblockN{Changsheng Chen \hspace{3cm} Wai Ho Mow}
\IEEEauthorblockA{\emph{HKUST Barcode Group}\\
School of Electrical and Computer Engineering\\
The Hong Kong University of Science and Technology\\
Email: \{eecschen,eewhmow\}@ust.hk}}


%


\maketitle

\begin{abstract}
Display-camera communication has become a promising direction in both computer vision and wireless communication communities. However, the consistency of the channel measurement is an open issue since precise calibration of the experimental setting has not been fully studied in the literatures. This paper focuses on establishing a scheme for precise calibration of the display-camera channel performance. To guarantee high consistency of the experiment, we propose an accurate measurement scheme for the geometric parameters, and identify some unstable channel factors, e.g., Moire effect, rolling shutter effect, blocking artifacts, inconsistency in auto-focus, trembling and vibration. In the experiment, we first define the consistency criteria according to the error-prone region in bit error rate (BER) plots of the channel measurements. It is demonstrated that the consistency of the experimental result can be improved by the proposed precise calibration scheme.
\end{abstract}

\IEEEpeerreviewmaketitle

\section{Introduction}

Display-camera communication has recently gained significant attention due to the pervasive and advancing of mobile phone camera \cite{Hranilovic,Perli2000,Hao2012,Wang2014,Ashok2014}. In the channel transmitter, the message is encoded and modulated into an image frame, e.g., a barcode image, and show on a conventional display. At the receiving end, a mobile camera serves as the receiver. The advantages of such display-camera channel are non-trivial. Firstly, the communication requires no extra hardware module except a pair of camera and display which are available on almost each off-the-shelf mobile phone. Secondly, compared with existing short range wireless communication technology, it requires no transmissions in the ever congested spectrum band and therefore creates no inferences with other devices. Last but not least, the security and privacy during communication can be well controlled by adjusting the visible distance and direction \cite{Hao2012, Wang2014}. The potential applications of the display-camera communications includes, but not limit to, information retrieval in a shopping mall from a large display, teaching material distribution in a classroom from the projector, multimedia file sharing from phone to phone between different users. The advantages are obvious and the potential applications are promising.

Researchers have been working on improving the channel reliability and throughput. Very few attentions have been paid to the experimental setup of the display-camera channel. However, the experimental parameters calibration is not a trivial problem. Firstly, it is not an easy task to set the experimental parameters accurately. In particular, it is difficult to calibrate the capturing angles between display and camera with conventional measurement tools, e.g., ruler and protractor. What's worse, it is found that the channel performance is sensitive to the experimental setting especially when high channel throughput is needed. As is shown in Section~\ref{sec:Experiment}, the channel performance has changed significantly even though the experimental setting deviates 2 degree in angle and 3 cm in the distance. In the literatures, some works focus on the case without perspective distortion \cite{Hranilovic, Wang2014} which is an ideal case in the practical applications; while others consider a wide range of capturing angles but no details on the measurement are given \cite{Perli2000,Hao2012,Ashok2014}.

The imprecise description of the experiment setting and the sensitivity of the channel performance lead to the issue of low consistency of the channel measurement results. However, consistency is the fundamental requirement of scientific experiments since the demonstration of experimental result shouldn't be based on a single event \cite{Sprott2000}. In this paper, we aim at addressing the issue of consistency of the display-camera channel measurements. The paper is organized as follows. Section \ref{sec:Proposed} describes our proposed channel calibration scheme. Section \ref{sec:Experiment} shows the consistencies of the channel measurement results with or without considering the proposed calibration method. Section \ref{sec:Conclusion} concludes this paper.

\section{The Proposed Channel Calibration Scheme}
\label{sec:Proposed}

In this section, some detail treatments of the experimental setup are proposed to guarantee the consistency of the experiment results. The contributions are mainly in two parts:
\begin{enumerate}[A.]
\setlength\itemsep{0.1cm}
\item \textbf{Ensure accurate geometric setups of the experiment}: To guarantee results with high consistency, the experimental parameters must be measured and set with ultimate care. We divide the parameters into two independent sets, i.e., the geometric and non-geometric parameters. The geometric ones include the display to camera distance, the capturing angle between the camera and display, the barcode resolutions/sizes on display and camera, etc. On the other hand, ambient light intensity, display brightness, image blurriness are classified as the non-geometric parameters which describes the transmitted signal energy. In this paper, we focus on the geometric parameters since the experimental results are more sensitive to the geometric parameters than the non-geometric ones. Especially, the capturing angle and the captured barcode size affect the channel performance heavily \cite{Ashok2014} due to the change of received signal energy.
\item \textbf{Avoid unstable channel states}: An experiment with certain experiment setting (experiment state point) is considered as consistent only when ``there is an open set around that point within which the result of that experiment is the same'' \cite{Written1980,Campbell1982}. In each setup of the display-camera communication channel, it presences one or multiple dominant error factors. The dominant factors can be a specific process of the decoding pipeline. For example, the barcode corner detection, image binarization, symbol synchronization, etc. It is important to identify these factors and ensure that they are not changed in another instance of the experiment setup. Therefore, it is crucial to identify the dominant error sources in the channel and avoid conducting experiments in the unstable states where the experimental results are sensitive to tiny deviation of the setup.
\end{enumerate}

In the following subsections, details on the above mentioned two aspects will be given.

\subsection{Precise Setup of the Geometric parameters}
\label{subsec:Geometric}

\begin{figure}
\centerline{
\hspace{0.08cm}
\begin{minipage}[c]{.45\linewidth}
  \centering
  \centerline{\includegraphics[width=3.5cm]{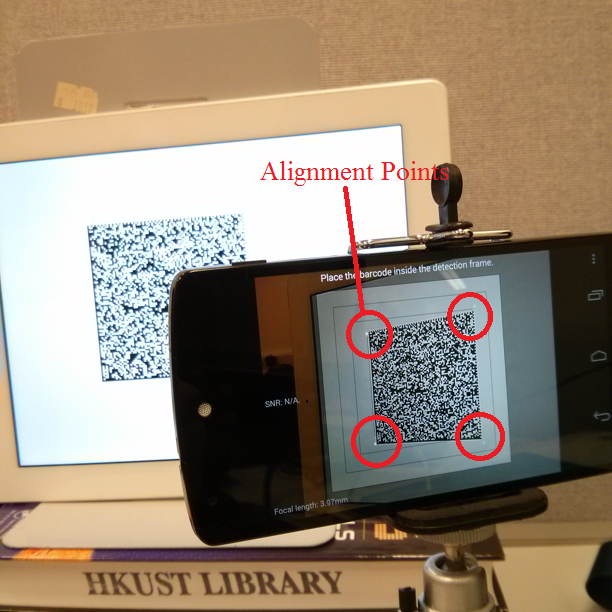}}
  \centerline{(a)}\medskip
\end{minipage}
\hspace{0.25cm}
\begin{minipage}[c]{.45\linewidth}
  \centering
  \centerline{\includegraphics[width=3.5cm]{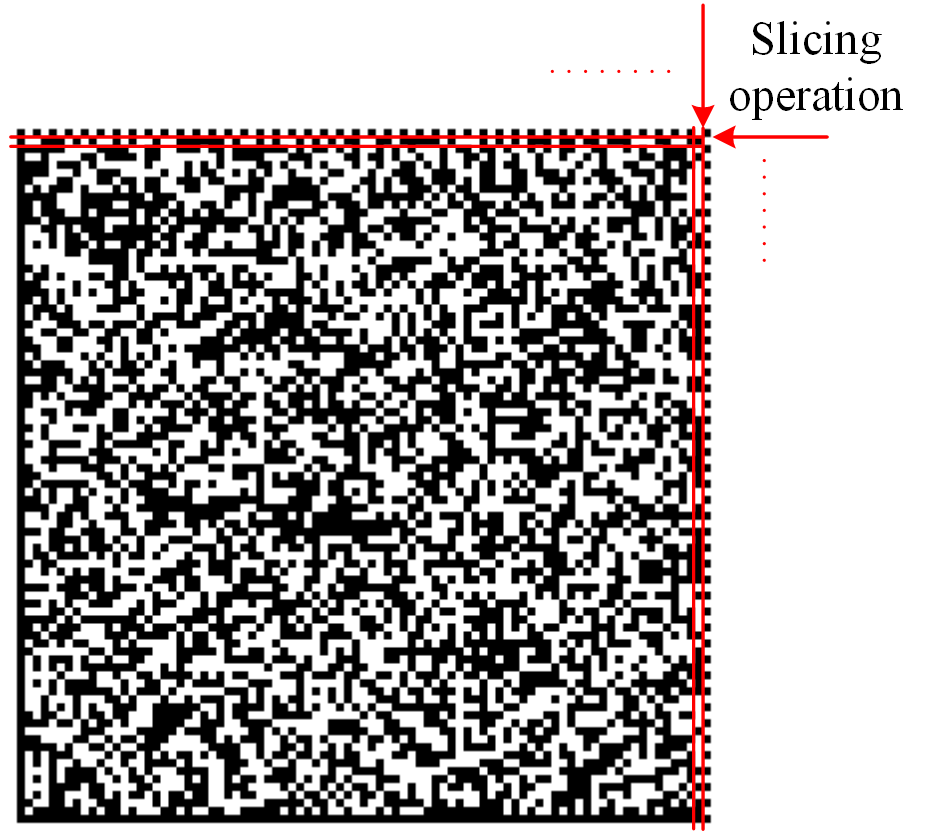}}
  \centerline{(b)}\medskip
\end{minipage}
}
\caption{The experimental setup. (a) A demo with 20 degree capturing angle using Nexus 5 and the new iPad. The alignment points are circled. (b) The barcode pattern used in our experiment and the illustration of slicing operation.}
\label{fig:ExpDemo}
\end{figure}

\begin{figure}
\centerline{
\includegraphics[width=7.5cm]{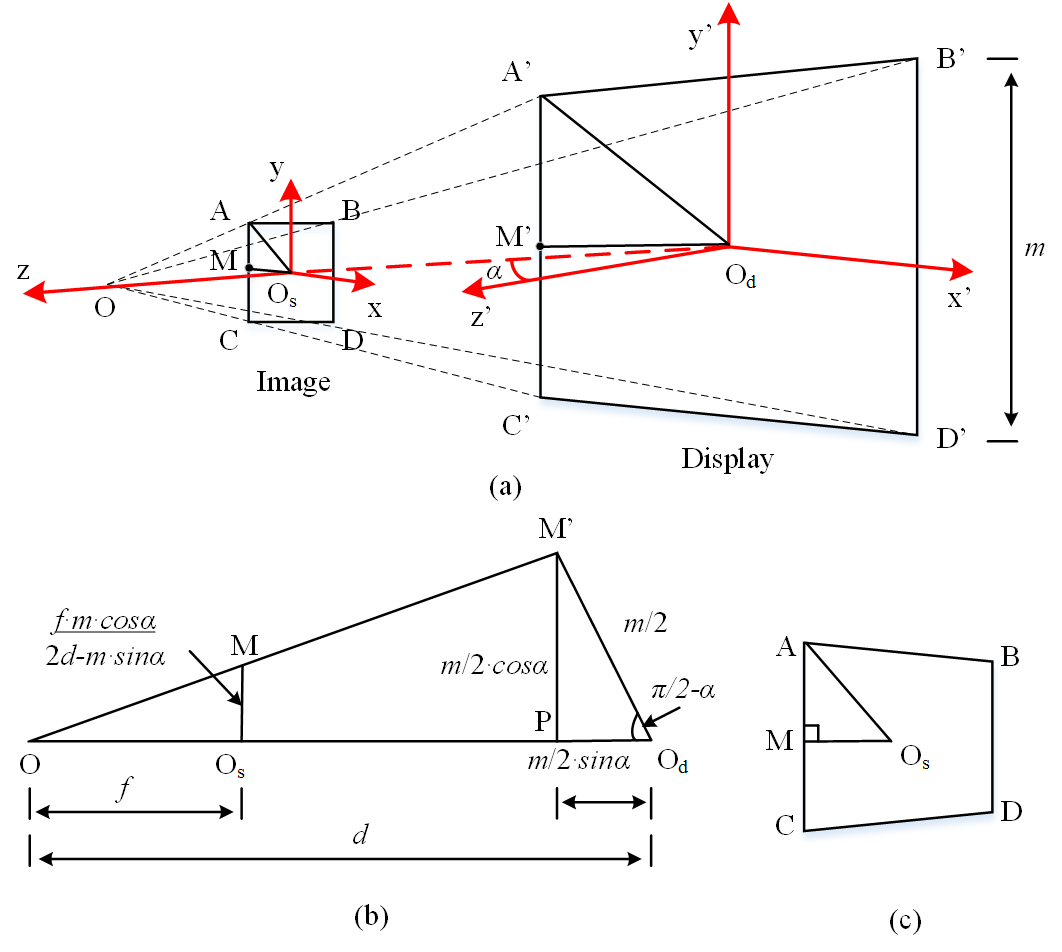}
}
\caption{The Pin-hole camera geometry model: (a) The mapping from display plane ($x'y'$) to image plane ($xy$); (b) Illustrate the calculation of $AO_c$ from $A'O_d$ with the principle of similar triangles; (c) the captured barcode on the image plane.}
\label{fig:DisplayCamera}
\end{figure}

To have a precise setup of the geometric parameters, we propose to use four reference points to align with the barcode corners as illustrated in Fig.~\ref{fig:ExpDemo}~(a). Given a fix displayed barcode size, the four points are set such that the aligned barcode is with the desirable capturing angle, distance and image resolutionp. Once the four barcode corners are aligned with the reference points and the displayed barcode size is fix, all the geometric parameters are set accurately. The problem of setting the geometric parameters accurately has been simplified to alignment of the four corners. The calculations of the four alignment points are explained with the display-camera model illustrated in Fig.~\ref{fig:DisplayCamera}.

In the Pin-hole camera model illustrated in Fig.~\ref{fig:DisplayCamera}~(a), the displayed barcode $A'B'C'D'$ on display plane $x'y'$is mapped to the captured barcode $ABCD$ on camera sensor plane $xy$. The camera principle axis $z$ passes through the optical center $O$, the captured barcode center $O_c$ and the displayed barcode center $O_d$. The distance between the displayed barcode center to the optical center $OO_d$ is denoted as $d$, while $f$ is the focal length parameter which is fixed in most off-the-shelf phone cameras \cite{iPhone2014,Nexus2014}. In this model, for the ease of experimental setup, the capturing angle $\alpha$ is set as with one degree of freedom, i.e., $z$ is co-plane with $x'z'$ and the angle between $z$ and $z'$ is $\alpha$. For experiments which study multiple degree of freedoms, the angle with other planes can always be added.

Let the displayed barcode size be $m\times m$, and $M, M'$ in Fig.~\ref{fig:DisplayCamera}~(a) be the center of $AB$ and $A'B'$, respectively. Based on the principle of similar triangle for $\Delta_{OMO_s}$ and $\Delta_{OMP}$ illustrated in Fig.~\ref{fig:DisplayCamera}~(b), it can be calculated that
\begin{align}
|MO_s|=\frac{f\cdot m\cdot \cos \alpha}{2d-m\cdot \sin\alpha}.
\end{align}
\noindent On the other hand, triangles $\Delta_{OAM}$ and $\Delta_{OA'M'}$ are also similar triangles, i.e.,
\begin{align}
\frac{|AM|}{|AM'|} = \frac{|MO|}{|M'O|} = \frac{|MO_s|}{|M'P|}.
\end{align}
\noindent Given that $|AM'| = m$, $|AM|$ can be computed as
\begin{align}
|AM| = \frac{f\cdot m}{2d-m\cdot \sin\alpha}
\end{align}
\noindent With $|MO_s|$ and $|AM|$, the coordinate of point $A$ w.r.t the origin of the image sensor $O_s$ can be calculated, so as the coordinates for other three corners. The locations of the four corners can be computed as:
\begin{align}
A=(-\frac{f\cdot m\cdot \cos \alpha}{2d-m\cdot \sin\alpha}, \frac{f\cdot m}{2d-m\cdot \sin\alpha}), \nonumber \\
B=(\frac{f\cdot m\cdot \cos \alpha}{2d+m\cdot \sin\alpha}, \frac{f\cdot m}{2d+m\cdot \sin\alpha}), \nonumber \\
C=(-\frac{f\cdot m\cdot \cos \alpha}{2d-m\cdot \sin\alpha}, -\frac{f\cdot m}{2d-m\cdot \sin\alpha}), \nonumber \\
D=(\frac{f\cdot m\cdot \cos \alpha}{2d+m\cdot \sin\alpha}, -\frac{f\cdot m}{2d+m\cdot \sin\alpha}) \nonumber
\end{align}
It should be noted that the above location in length should be converted to image coordinates with the knowledge of pixel size on the imaging sensor. During the experiment, four corner locations will be calculated according to the predetermined parameters and the above geometric model. The computed coordinates are marked in the preview window on the mobile App. Therefore, the problem of setting the geometric parameters accurately has been simplified to alignment of the four barcode corners to the reference points.

\subsection{Avoid the Unstable Experiment States}

In this part, we discuss several undesirable factors of the experiment setup which leads the unstable experiment states. The settings come from a wide range of factors including the trembling of camera, the modulation scheme of LED display, the auto focus operation of camera, the sampling of the displayed pixels with the camera sensor, and the effect of tiny vibrations. In the following, detail discussions will be presented for these factors.

\subsubsection{Avoid rolling shutter effect from the CMOS sensor}

Rolling shutter scheme is a popular data acquisition scheme for the digital camera with CMOS sensor. It reads out the imaging data from the CMOS sensor pixels row by row sequentially from top to bottom \cite{Bigas2006}. This is a common scheme in the state-of-the-art high resolution CMOS sensor to enable the sensor to continuously gather photons during the exposure process and thus increase the sensitivity of the imaging sensor. The major disadvantage for the rolling shutter scheme is that the time difference between retrieving the pixel data from the top and bottom of the sensor introduces time delay and therefore the rolling shutter effect \cite{Nakamura2005}. Thus, it is not appropriate to be used in capturing the scene with fast-changing environment.

Unfortunately, the display-camera channel can be viewed as a periodic fast-changing channel over space due to the Pulse-Width Modulation (PWM) scheme which are popularly used in the conventional displays \cite{Hainich2011}. PWM scheme is used to control the brightness or the backlight intensity of a screen. Due to the binary nature of the display back light, a pixel can only be turned on or off which corresponds to the maximum or minimum pixel intensities, respectively. To display gray level images, the PWM driver turns on the each pixel for a certain duration in each duty cycle to achieve the intermediate brightness levels. For a typical display driven by PWM scheme, the full duty cycle is about $5.5 \mbox{ ms}$ \cite{Hainich2011}. If the brightness level is set to 50\%, the pixel is turned on and off for the same duration, i.e., $~2.75 \mbox{ ms}$, in each duty cycle. On the other hand, the typical image exposure time for each video frame is from $1/200$ to $1/30$ s which means the time differences between retrieving the top and bottom row is much larger than the on-off durations \cite{Nakamura2005}. Therefore, several bright and dark bands correspond to the PWM on-off durations in each duty cycle can be observed in a single captured image due to the rolling shutter effect.

\begin{figure} [h!]
\centering{
\hspace{0.05cm}
\begin{minipage}[c]{.3\linewidth}
  \centering
  \centerline{\includegraphics[width=1in]{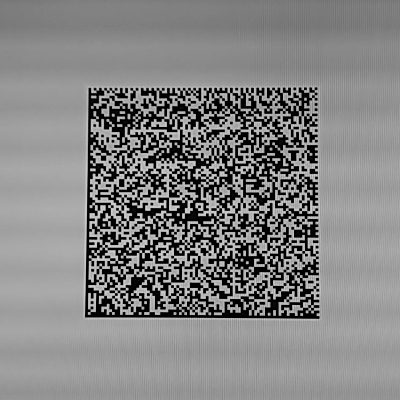}}
\end{minipage}
\hspace{0.05cm}
\begin{minipage}[c]{.3\linewidth}
  \centering
  \centerline{\includegraphics[width=1in]{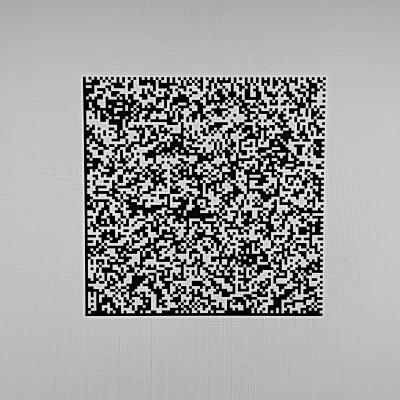}}
\end{minipage}
}
\caption{Barcode images with (left) and without (right) rolling shutter effect from the Dell E2313H display \cite{Dell2014} (driven by PWM scheme).}
\label{fig:RollingShutter}
\end{figure}

In order to remove the roller shutter effect on the display with PWM scheme, the brightness level should be tuned to full so that the transmitted signals is kept as static for each display pixel. As demonstrated in Fig.~\ref{fig:RollingShutter}, the rolling shutter effect with alternative dark and bright patterns can be observed clearly in the left image with 50\% display brightness; while it is eliminated with full brightness level and disable any automatic brightness adjustments. An alternative approach is avoid using the display with PWM scheme, a list of the possible choice of PWM free monitor can be found at \cite{Baker2014}.

\subsubsection{Avoid inconsistency from auto-focus operation}

Auto-focus refers to the camera operation which automatically adjust the distance from the lens and imaging sensor so that the capture image is visually sharp \cite{hirsch2012}. There are two types of auto-focus techniques, i.e., active and passive auto-focus. Active auto-focus measure the distance between the camera and the object of interest with an infrared beam and adjust the sensor-to-lens distance accordingly; while the passive one takes a series of video frames with varying sensor-to-lens distances and pick the best one by inspecting the image quality metric, e.g., contrast, in each image. Passive auto-focus is the dominating one in the state-of-the-art mobile phone cameras. Thus, it is of our main interest.

However, each trial of auto-focus may produce different results even if the geometric setting is fixed. This is mainly due to two reasons. First, the limited resources at hand, e.g., low quality of the image sequences, strict time constraint and limited computational power in the metric calculations. Second, the metric differences between the adjacent frames are usually very small due to the simple quality metric and the large depth-of-field of the mobile phone camera \cite{micheletti2010}. The slight difference in the sensor-to-lens distance across different images produce significant change of image quality in a single experiment. In order to maintain the consistency during the experiment, we suggest to turn off the auto-focus function after one trial at the beginning of each experiment, and the barcode image collection starts only after the auto-focus operation has finished.

In order to freeze the auto focus function, a modification to the auto focus loops was made. As shown in Algorithm~\ref{alg:autofocus}, an indicator \emph{counter} has been added to the generic auto-focus handling algorithm \cite{Zxing2014} to fix the focus setting after the camera focus is properly set.

\begin{algorithm}
\caption{The One Time Auto-Focus Algorithm}
\label{alg:autofocus}
\begin{algorithmic}[1]
\State Set auto-focus states: $\textit{counter} = 0$ and $\textit{focusing} = 0$;
\State Initialize preview mode parameters;
\BState \emph{loop}:
\If {$\textit{counter} = \textit{0}$}
\State $\textit{focusing} = 1$;
\State \textbf{goto} AUTO-FOCUS.
\State \textbf{close};
\EndIf\\

\Procedure{Auto-focus}{}
\If {Find best focus lens = SUCCESS}
\State $\textit{counter} = 1$; $\textit{focusing} = 0$;
\Else
\State $\textit{counter} = 0$; $\textit{focusing} = 0$;
\EndIf
\EndProcedure
\end{algorithmic}
\end{algorithm}

\subsubsection{Avoid Moire effect from the display}
\label{subsubsec:Moire}

It is well known that recapturing an images from the display without careful setup introduces the Moire effect. It is due to aliasing from sampling the display pixel grid with the camera sensor pixels \cite{Muammar2013}. The barcode detection performance is heavily affected by the Moire pattern since it introduces false barcode region. What's worse, the pattern is very sensitive to the geometric setting. A tiny shift in the camera position could lead to huge change of the Moire pattern thereafter affects the overall decoding performance. Researchers have been working on possible solutions to avoid the Moire effect in the recaptured image. Some suggestions are made to eliminate the Moire patterns, such as intentionally avoid sharp focus and preprocess the image with a frequency domain filter. However, these operations limit the sharpness of the barcode image and introduce inevitable loss into the images. Recently, Muammar and Dragotti \cite{Muammar2013} model the structure of the display pixel grid as 2D square form and show that the artifacts can be eliminated by simply setting the display-camera distance to a predetermined value. The distance can be calculated by the knowledge of camera focal length, the display and camera sensor pixel sizes, that is,
\begin{align}
\label{eq:Moire}
d_k = 2f\Big(\frac{kT_s}{T_d} + \frac{T_d}{4kT_s}+1\Big)
\end{align}
\noindent where $d_k$ is the desirable distance, $T_s, T_d$ are the pixel sizes on camera sensor and display, respectively, and $k$ is an arbitrary integer. With any integer $k$, the calculated distance $d_k$ can effectively eliminate the Moire pattern on the captured image. For a generic setting with the new iPad display and the Nexus 4/5 camera, we have $T_d=0.097 \mbox{ mm}$ (iPad), $T_s=1.1 \mbox{ $\mu$m}$ (Nexus 4) or $1.4 \mbox{ $\mu$m}$ (Nexus 5), and $f = 4.6 \mbox{ mm}$ (Nexus 4) or $4.0 \mbox{mm}$ (Nexus 5) \cite{Nexus2014, iPad2014}. The Moire pattern eliminating distances for both sets of equipment have been calculated in Table~\ref{tab:MoireDistance}. The four reference points mentioned in Section~\ref{subsec:Geometric} will be determined based on these Moire pattern eliminating distances. Therefore, Moire pattern in the captured image is a good indicator of whether the geometric setting is precise enough.

\begin{table} [h!]
\begin{center}
\caption{Capture distance with no Moire pattern (in meter).}
\label{tab:MoireDistance}
\begin{tabular}{ c | c | c | c | c | c }
  \backslashbox[30mm]{Equipment}{$k$}  & 1 & 2 & 3 & 4 & 5 \\ \hline \hline
  The new iPad with Nexus 4 & 0.21 & 0.11 & 0.077 & 0.060 & 0.050 \\ \hline
  The new iPad with Nexus 5 & 0.15 & 0.078 & 0.055 & 0.043 & 0.036
\end{tabular}
\end{center}
\end{table}

It should be noted that this scheme works only when the display and camera planes are parallel. However, it brings insightful clues for the choice of equipment which can alleviate the Moire pattern. As $\frac{T_d}{4kT_s}$ being the dominant terms in the bracket of Eq.~(\ref{eq:Moire}), a smaller display pixel size results in a smaller range for the Moire pattern eliminating distances which benefits different parts of the image with different but similar display to camera distance. In other words, even when the display and camera planes are not parallel, a display with higher PPI (Pixels Per Inch) produces less obvious Moire effect.

\subsubsection{Avoid Trembling and Vibration}

Display camera communication is pervasive and sometime ad hoc. The camera devices could be handheld with no stable station. This lead to the difficulty of producing consistent channel measure since the irregular hand trembles. This leads to unexpected motion blur of the captured images which lead to degradation of the image quality and the loss of parts of an image frame \cite{Chen2013} \cite{Wang2014}. In order to avoid abrupt change of the communication channels, it is suggested that a stationary platform with adjustable freedom in the viewing angle should be used to hold the camera. One example of such equipment is a mini tripod as shown in Fig.~\ref{fig:ExpDemo}~(a).

Besides the hand trembling, some tiny mechanical vibrations also causes the unexpected camera motion. Early studies \cite{Wulich1987} \cite{Rudoler1991} show that the mechanical vibration introduces inter-pixels interferences and limits the performances of an imaging sensor. The situations is worsen in our cases since a display is used as the transmitter. On the one hand, some inevitable Moire patterns are produced when the display and camera planes are not parallel. As discussed in Section~\ref{subsubsec:Moire}, the Moire effect is generated by sampling the display pixel grid with the camera sensor pixels. It is extremely sensitive to the vibration in the display camera channel. A small vibration of the camera or display leads to a big change of the Moire pattern which heavily affects the image preprocessing results, e.g., binarization output. On the other hand, the vibration causes tiny shifts of the barcode corners in the captured images. The slight difference in pixel or even sub-pixel level leads to very different results in the corner detection and symbol synchronization steps.

In the literatures, researchers have proposed several approaches to solve the issue caused by vibration. Siebert {\it et al.} \cite{Siebert2009} shows that the vibrations can be well estimated with a 3D homography setup using calibrated two cameras. In our setup, a pair of consecutive images can be considered as images from two independent camera. The vibration between the two image frames can then be estimated using the same homography model and corresponding compensation can be carried out. The advanced mobile phones offers image stabilization functionality in optical and software level. For example, the iPhone 6 Plus \cite{iPhone6P2014} employs optical image stabilization which takes advantages of gyroscope and the motion sensors to measure motion data and provide accurate information on the lens movement. The motion blur produced by hand shake can then be compensated even under lower light condition.

Therefore, the experiment setup is suggested to be stationed at a desk without any vibration sources, such as, computer cooling system, hard disk motor, etc. A tablet which has no moving components is very suitable display for our experiment, and the phone models with image stabilizer for the camera is preferable.

\section{Experimental Results}
\label{sec:Experiment}

In this section, the experimental procedure is firstly introduced and the experimental results with and without considering the proposed channel calibration scheme are shown.

\subsection{Introduction to Experimental Procedure}
\label{subsec:RepeatMetric}

The purpose of the display-camera channel calibration experiment is to identify the range of error-prone region since it is important in error correction code with unequal error protection (UEP) \cite{Goldsmith2005} and other advanced coding schemes. In this part, we first describe our experimental setting to produce consistent results and followed by a methodology to analyze the experimental results.

\subsubsection{Experimental setups}

Given the considerations discussed in Section~\ref{sec:Proposed}, the experimental parameters are set as follows:

\begin{itemize}
  \item \emph{Display}: the new iPad with Retina display ($2048 \times 1536$ pixels);
  \vspace{-0.15cm}
  \item \emph{Camera}: Nexus 4/5 run in VGA ($640 \times 480$ pixels) preview mode;
  \vspace{-0.15cm}
  \item \emph{Camera stand}: A mini tripod with adjustable orientation and height;
  \vspace{-0.15cm}
  \item \emph{Barcode pattern}: a generic binary barcode with datamatrix-like structure as shown in Fig.~\ref{fig:ExpDemo}~(b).
  \vspace{-0.15cm}
  \item \emph{Barcode size}: $10 \times 10 \mbox{ cm}^2$ on display;
  \vspace{-0.15cm}
  \item \emph{Barcode dimension}: $87 \times 87 \mbox{ module}^2$;
  \vspace{-0.15cm}
  \item \emph{Distance}: 21 and 15 cm from the camera to barcode center for Nexus 4 and 5, respectively, according to Eq.~(\ref{eq:Moire});
  \vspace{-0.15cm}
  \item \emph{Capture angles}: -20, 0, and 20 degree.
  \vspace{-0.15cm}
  \item \emph{Brightness}: 250-350 lx as measured by a lux meter.
\end{itemize}

It should be noted that our barcode pattern is not exactly the same as the datamatrix code \cite{Datamatrix2006}. The main difference is that our barcode pattern has an odd number of modules in each dimension (e.g., 87 in our setup) while the datamatrix code has an even number of modules. This has been modified to improve the detection accuracy of the top right corner. However, our scheme of the channel calibration does not limit to a specific barcode pattern. For the barcode dimension, it is chosen since it demonstrates the high throughput ($85\times 85$ bits/frame) performance which is of our main interest. The display-camera distance is fine-tuned to minimize the effect of Moire pattern according to Eq.~(\ref{eq:Moire}). The setting of capture angles covers a wide range of practical applications. For each angle setting, we pre-compute a set of four reference points as shown in Section~\ref{subsec:Geometric}. The user are required to align the reference points with the barcode corners. A simple demo of the 20 degree experimental setting are shown in Fig.~\ref{fig:ExpDemo}~(a).

In each set of experiment, 500 images are collected to achieve high reliability, i.e., a good enough confidence interval. The 95\% confidence interval can be computed by \cite{Belia2005}:
\begin{align}
\label{eq:ConfidenceInterval}
I_C = 2\sqrt{p\cdot(1-p)}/\sqrt{N_I},
\end{align}
\noindent where $p$ is the BER for a given module and $N_I$ is the number of images, i.e., 500 for our experiment. It means the actual BER has 95\% probability to be in the interval $[ p-I_C, p+I_C]$. For $p=0.1$, $I_C = 0.026$ and the 95\% confidence interval $[0.074, 0.126]$. In other words, with 500 images and 0.1 BER, the experiment result has achieved 95\% confidence interval in $\pm 0.026$ interval. The 95\% confidence interval for $p=0.2,0.3,0.4$ can also be computed as $\pm 0.036, \pm 0.042, \pm 0.044$.

\subsubsection{Result Analysis: Consistency Measure}

\begin{figure}
\centerline{
\includegraphics[width=8cm]{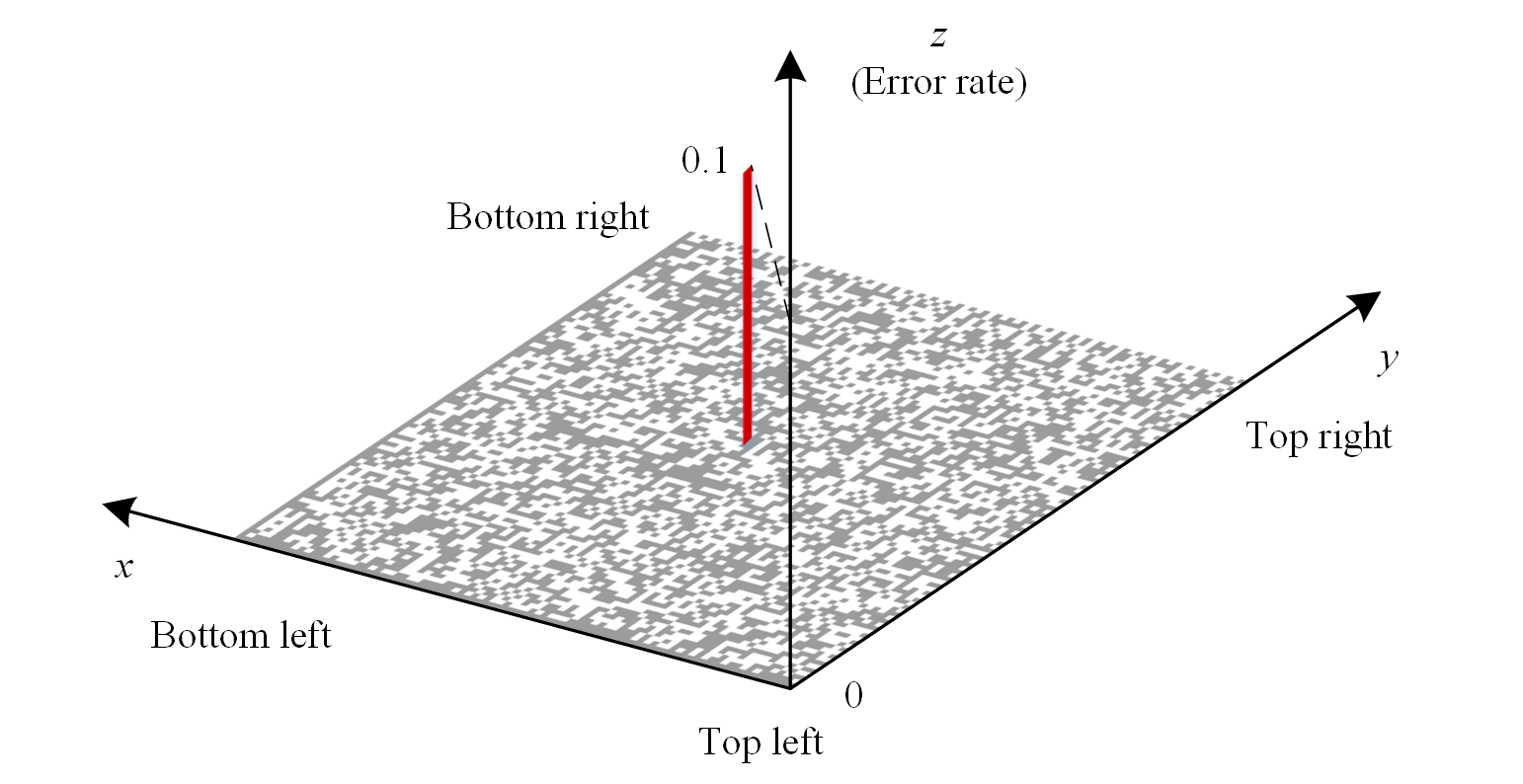}
}
\caption{An illustration of the bit error rate plot over the barcode region.}
\label{fig:ErrPlot}
\end{figure}

After collecting the barcode images, they are forwarded to the decoder and the demodulation bit error rate (BER) plot is generated by assembling the error probability of each modules according to the module position. As illustrated in Fig.~\ref{fig:ErrPlot}, the error plot is overlaid with the barcode image to shown the position correspondences. The BER value at a given position in the xy-coordinate shows the performance of the corresponding barcode module. In this toy example, an error-prone module with 0.1 error rate is found at the center of the barcode region.

Across different experiments, the BER plots are compared to evaluate the consistency of the experiments. In this study, the locations of the error-prone region rather than the amplitudes of individual peaks are of our main interest since the error-prone locations are the primary concerns of some advanced error protection schemes and channel studies. Therefore, preprocessing of the BER plots is needed to eliminate the noisy spikes before evaluating the consistency. One generic approach is that a thresholding operation is applied on the BER plots so that only the dominant peaks are kept, and the region within $D$ modules L1-distance to the dominant peaks are labeled as the error-prone region, $P_1$ and $P_2$. In the experiment, the threshold is empirically selected to be 50\% percent of the mean amplitude of the top 10 dominant peaks and $D$ is set to be 3 module. The common error-prone regions in both BER plots are obtained by $P_1 \bigcap P_2$, while the overall error-prone region are $P_1 \bigcup P_2$. $E_{c}$ and $E_{a}$ denotes the number of nonzero modules for the intersection and union sets, respectively. Finally, the consistency metric is defined as:
\begin{align}
\label{eq:RepeatabilityScores}
R = E_{c}/E_{a}
\end{align}
\noindent The higher the metric, the more consistent the two experiments are.

\subsection{Consistency Experiment: with the Proposed Scheme}
\label{subsec:Repeatable}

\begin{figure*} [t!]
\centerline{
\includegraphics[width=17cm]{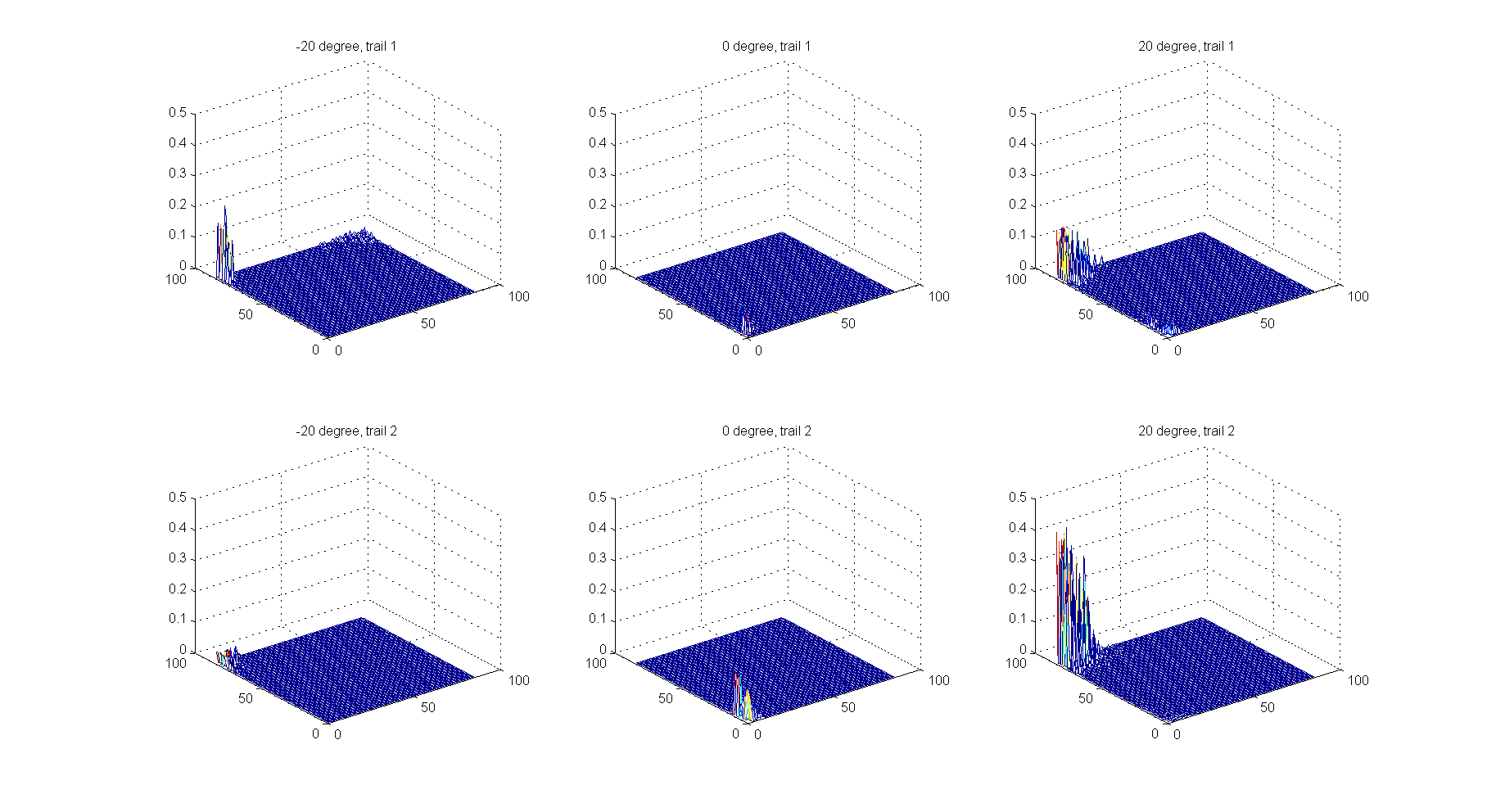}
}
\caption{The BER plots for our experimental results at -20, 0 and 20 degree capturing angle with Nexus 5 camera and the new iPad display. The consistency metrics for the three columns are: 0.56, 0.40 and 0.77 for the -20, 0 and 20 degree conditions, respectively.}
\label{fig:ErrDistsN5}
\end{figure*}

In this part, we demonstrate that the consistencies of the experiments can be maintained by accurately setting the geometric with the proposed scheme. As shown in Fig.~\ref{fig:ErrDistsN5}, the consistency metrics for the above distributions are computed as 0.56 for -20 degree, 0.39 for the 0 degree, and 0.77 for 20 degree. It can be seen that the top left and bottom left corner regions are the error-prone positions in the error plots regardless of the capturing angles. This is due to the datamatrix-like finder and timing patterns design. As can be seen in Fig.~\ref{fig:ExpDemo}~(b), the top and right edges of the barcode consist of alternative black and white modules which are used in the symbol synchronization. Errors in locating the barcode position heavily affects the symbol synchronization accuracy. This is due to the red slicing lines for symbol alignment are drawn from top to bottom or from right to left. The top left, bottom left and bottom right corner regions suffer the most from the symbol synchronization errors.

\begin{figure}
\centering{
\begin{minipage}[c]{.45\linewidth}
  \centering
  \centerline{\includegraphics[width=1.25in]{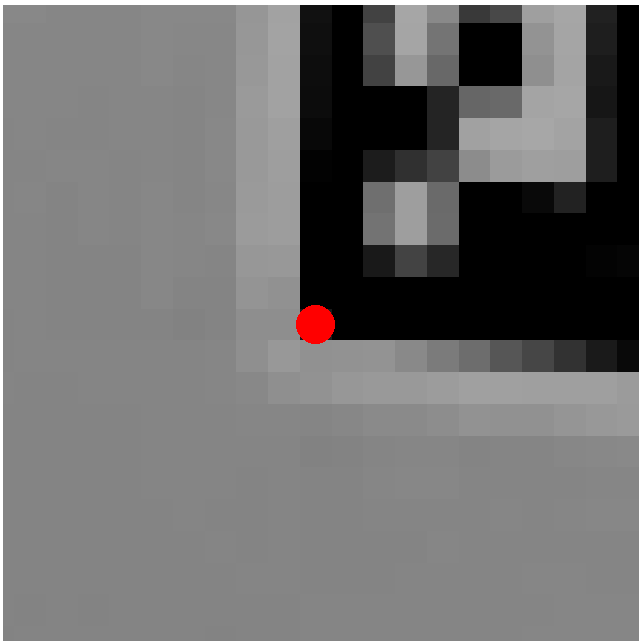}}
\end{minipage}
\hspace{0.25cm}
\begin{minipage}[c]{.45\linewidth}
  \centering
  \centerline{\includegraphics[width=1.25in]{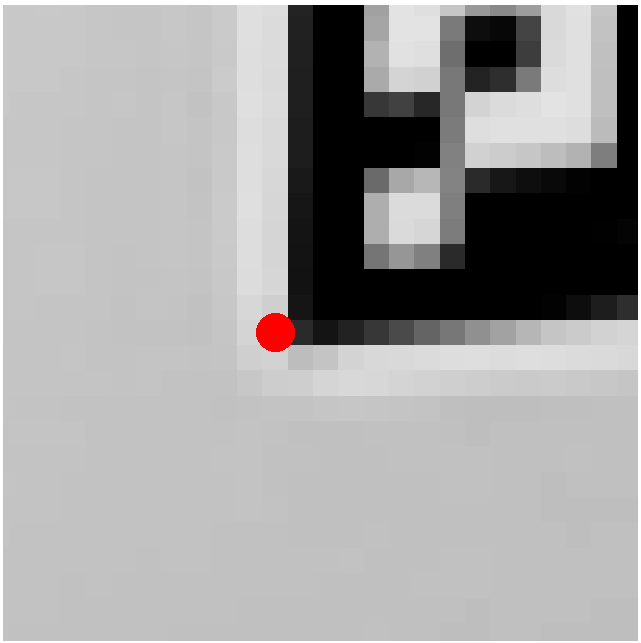}}
\end{minipage}
\\
\vspace{0.25cm}
\hspace{0.08cm}
\begin{minipage}[c]{.45\linewidth}
  \centering
  \centerline{\includegraphics[width=1.25in]{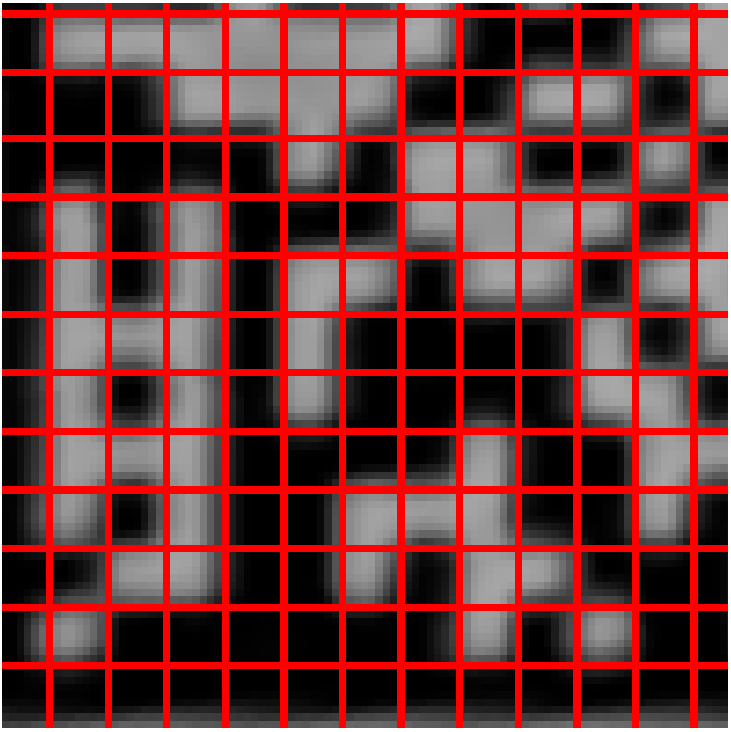}}
  \centerline{(a) first set}\medskip
\end{minipage}
\hspace{0.25cm}
\begin{minipage}[c]{.45\linewidth}
  \centering
  \centerline{\includegraphics[width=1.25in]{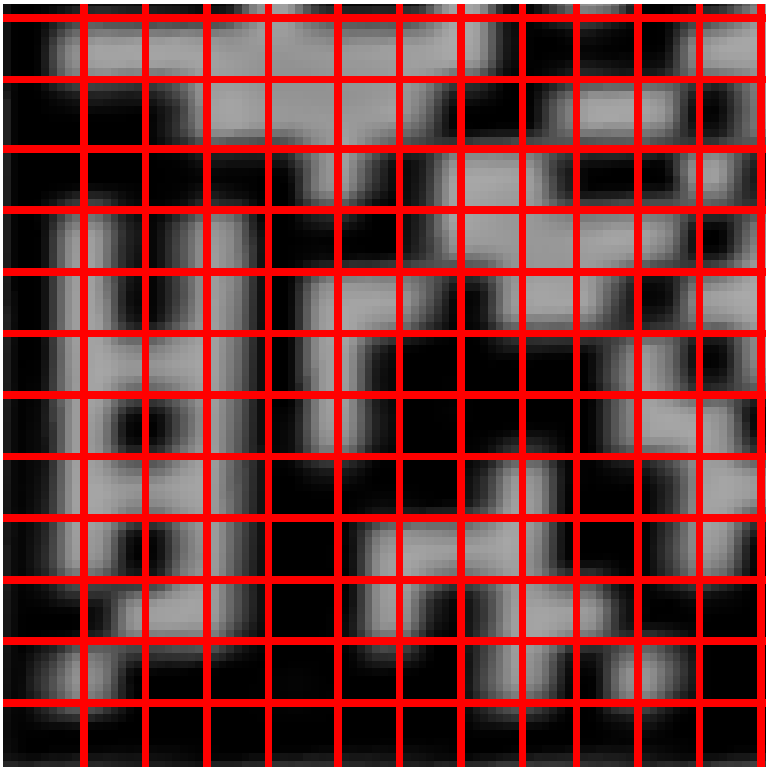}}
  \centerline{(b) second set}\medskip
\end{minipage}
}
\caption{The corner detection and symbol synchronization results for both experiments in the case of 20 degree capturing angle. Top: the detected bottom left corners marked with red dots; Bottom: the corresponding symbol synchronization performances at the bottom left region.}
\label{fig:ErrAnalysis}
\end{figure}

It is also worth mentioning that the differences of the amplitude in the two corresponding error plots are due to small alignment differences in the two experimental setting. As can be seen in Fig.~\ref{fig:ErrAnalysis}, the first set of images has higher detection accuracy in the bottom left corner than that of the second ones. Thus, smaller symbol synchronization error and lower amplitudes in the BER plot are obtained for the first set of images.

The results for both Nexus 4 and Nexus 5 are summarized in Table~\ref{tab:Repeatability}. It is understandable that when the channel distortions are strong and deterministic, the experiments should have high consistency. For example, under -20 and 20 degree perspective distortion, the consistency scores are generally higher than those of experiments at 0 degree.

\begin{table} [h!]
\begin{center}
\caption{The Consistency Scores.}
\label{tab:Repeatability}
\begin{tabular}{ c | c | c | c }
  \backslashbox[30mm]{Equipment}{Angle (in degree)}  & -20 & 0 & 20 \\ \hline \hline
  The new iPad with Nexus 4 & 0.68 & 0.52 & 0.82 \\ \hline
  The new iPad with Nexus 5 & 0.56 & 0.40 & 0.77
\end{tabular}
\end{center}
\end{table}

\subsection{Inconsistent Experiments: without the Proposed Scheme}
\label{subsec:Unrepeatable}

Slight deviations in the geometric parameters could lead to very different decoding performance and therefore a very different error plot. Without precise calibration of the geometric parameters, the consistency of experimental results from independently collected images can not be guaranteed. In this part, we show that a small inaccuracy in setting the angle and distance lead to totally different decoding performance.

\begin{figure}
\centering{
\begin{minipage}[c]{.45\linewidth}
  \centering
  \centerline{\includegraphics[width=1.8in]{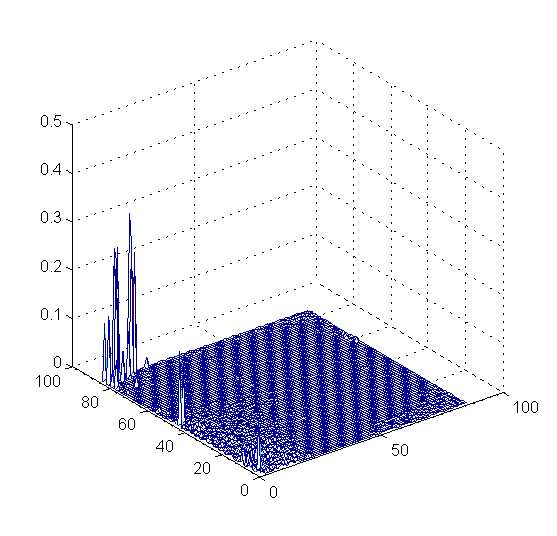}}
  \centerline{(a)}\medskip
\end{minipage}
\hspace{0.25cm}
\begin{minipage}[c]{.45\linewidth}
  \centering
  \centerline{\includegraphics[width=1.8in]{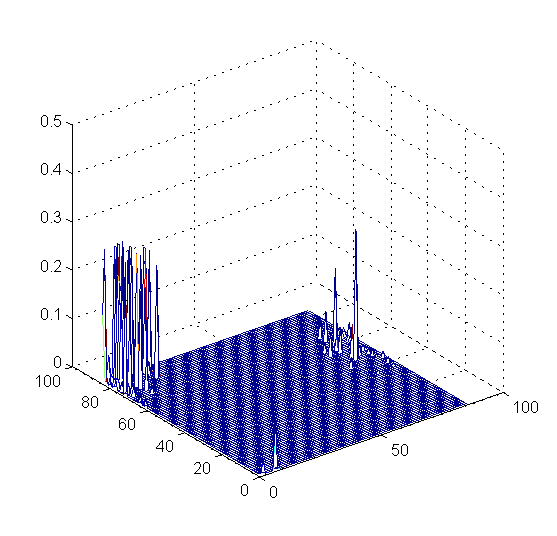}}
  \centerline{(b)}\medskip
\end{minipage}

\begin{minipage}[c]{.5\linewidth}
  \centering
  \centerline{\includegraphics[width=2in]{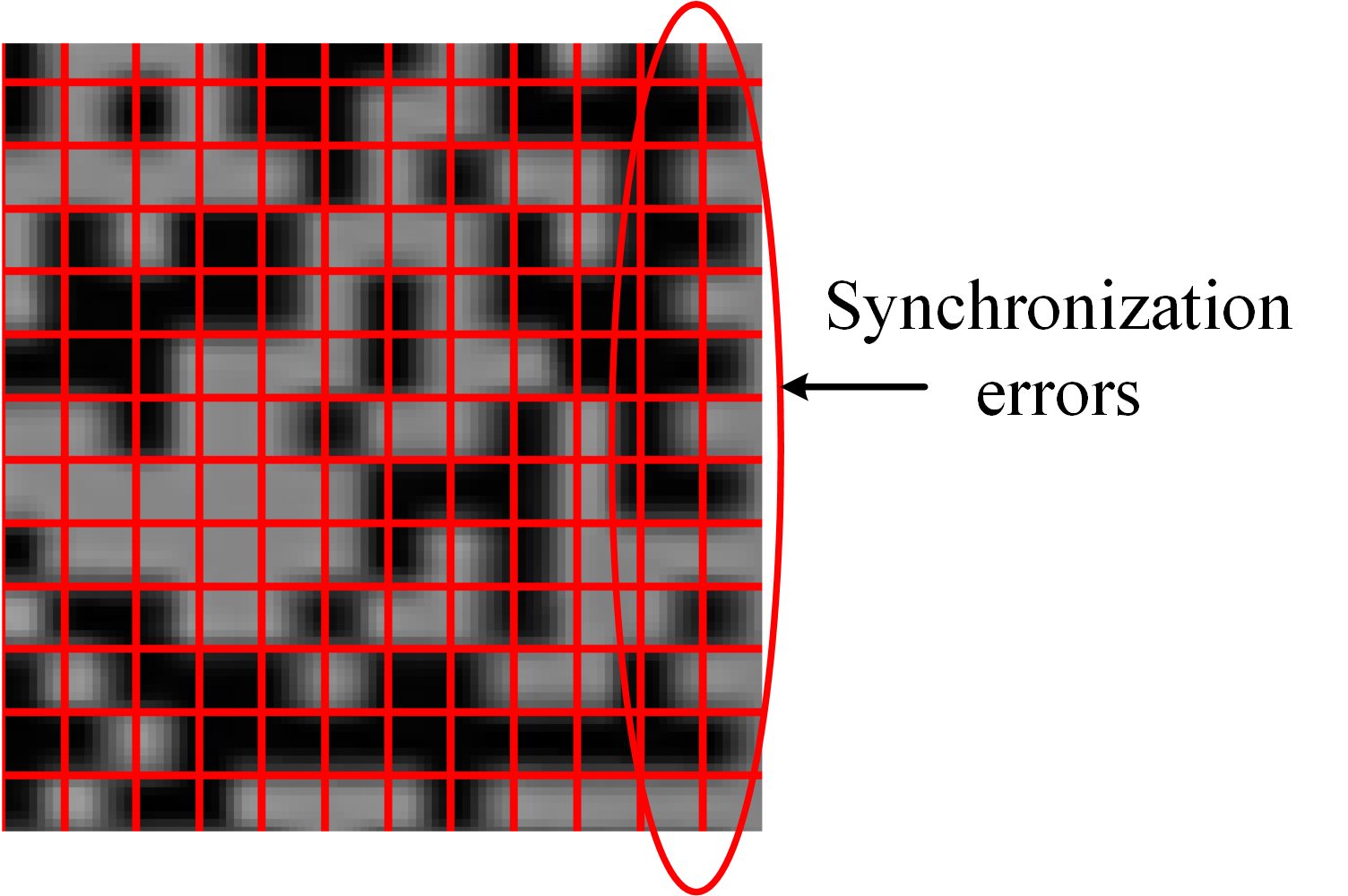}}
  \centerline{(c)}\medskip
\end{minipage}
}
\caption{Error plots of the experimental results by Nexus 4 and the new iPad with: (a) precise geometric setting, i.e., -20 degree and 21 cm; (b) -22 degree and 18 cm. (c) The synchronization error due to lens distortions in the right edge of the barcode.}
\label{fig:Nonrepeatable}
\end{figure}

One set of experimental images are collected with -22 degree angle and 18 cm distance using Nexus 4 and the new iPad. Such deviations can easily be made if the angle and distance measurements are not done properly. For example, measuring the distance when the phone and display are not properly aligned ($O, O_s$ and $O_d$ do not lie on the same axis as shown in Fig.~\ref{fig:DisplayCamera}), and roughly setting the angle with a pair of protractor and rule. It turns out that the consistency score drop significantly when compared to the error plots with precise geometric setups. As shown in Fig.~\ref{fig:Nonrepeatable}~(a) and (b), the position of error-prone region has changed significantly and the consistency score has dropped to 0.22 though the geometric setup only deviates 2 degree in viewing angle and 3 cm capturing distances. The error-prone region near the right edge is due to the lens distortion gets more and more serious when the viewing angle increases and the capturing distance decreases as shown in Fig.~\ref{fig:Nonrepeatable}~(c).

\section{Conclusion}
\label{sec:Conclusion}

In this paper, we have established a precise calibration scheme for the geometric parameters in the display-camera communication channel. Four reference points have been pre-computed according to the predetermined geometric parameters. The problem of accurate setup of the parameters has been simplified to alignment of four corner with the four reference points. Careful settings of the equipment and experimental parameters are also needed to avoid some unstable channel states, such as, Moire effect, rolling shutter effect, blocking artifacts, auto-focus inconsistency, trembling and vibrations. In the experiment, the BER error plots of the captured barcode images are analyzed and the consistency criteria is defined. We demonstrate that setting the geometric parameter accurately and avoiding the unstable factors with the proposed scheme can improve the consistency of the experiment significantly.


\section*{Acknowledgment}

This work was supported by the Hong Kong Research Grants Council (Project Number 616512). The authors would like to acknowledge the other members of the HKUST Barcode Group for making various contributions on which this work is based.

\bibliographystyle{unsrt}
\bibliography{SystematicExperiment}  

\end{document}